# A METHODOLOGY FOR MULTISENSORY PRODUCT EXPERIENCE DESIGN USING CROSS-MODAL EFFECT: A CASE OF SLR CAMERA

**Takuma Maki and Hideyoshi Yanagisawa**
The University of Tokyo
hide@mech.t.u-tokyo.ac.jp

## ABSTRACT

Throughout the course of product experience, a user employs multiple senses, including vision, hearing, and touch. Previous cross-modal studies have shown that multiple senses interact with each other and change perceptions. In this paper, we propose a methodology for designing multisensory product experiences by applying cross-modal effect to simultaneous stimuli. In this methodology, we first obtain a model of the comprehensive cognitive structure of user's multisensory experience by applying Kansei modeling methodology and extract opportunities of cross-modal effect from the structure. Second, we conduct experiments on these cross-modal effects and formulate them by obtaining a regression curve through analysis. Finally, we find solutions to improve the product sensory experience from the regression model of the target cross-modal effects. We demonstrated the validity of the methodology with SLR cameras as a case study, which is a typical product with multisensory perceptions.

## 1  INTRODUCTION

Users perceive a product through multisensory modalities such as vision, hearing, and touch; these modalities often interfere with each other. Studies have been conducted on these multimodal interactions from different viewpoints. In the field of psychology, Charpentier (1891) discovered that the visual size of an object alters its weight perception, which is well known as size-weight illusion. Driver and Spence (1998) have revealed that perceptions of spatial location by vision, audition, and touch are correlated. In the field of Food Science, Deliza and MacFie (1996) investigated the effects of visual expectations with regard to food and its actual taste, while Schifferstein et al. (2013) showed how the visual design of a package affects consumer's food experience.

In this paper, we term multisensory interactions between multiple modalities *cross-modal effects*. Many studies have shown the potential ability of cross-modal effects to improve products' sensory experiences. For example, Ludden et al. (2009) examined and demonstrated the effects of visual-tactile incongruity on surprise reactions with products that were similar in visual appearance, but differed in tactile characteristics. Also, Lederman, Thorne, and Jones (1986) noted that perception of texture is correlated to the weighted linear combination of the information from vision and touch. These facts suggest products can be designed to evoke delightful surprise by manipulating the properties of vision and touch modalities.

As this example shows, a cross-modal effect can be applied to products' sensory experiences. For another instance, using size-weight illusion it is possible to reduce the perceptive weight by making the size bigger to make users perceive the weight lighter than they expected without reducing the actual weight. Therefore, even if physical limitations exist, the cross-modal effect enables to design sensory product experiences that we could not design within a single modality and exceed the limitations at perceptual level. Thus, we considered that a cross-modal effect could be applied to product experience design.

Yanagisawa, Miyazaki, and Bouchard (2017) proposed Kansei modeling methodology to extract users' comprehensive cognitive structure of product experience and also showed a way to find cross-modal effect opportunities and how to formulate the found cross-modal effect with expectation theory (Yanagisawa, Miyazaki, & Nakano, 2016; Yanagisawa & Takatsuji, 2015). By applying the methodology, they found that the visual size of a product (e.g., a hair dryer) affected the perception of the sound (e.g., the exhaust sound), which led them to a sensory design solution. This showed that a sensory experience designed to have a cross-modal effect could affect a user's feeling about the product and improve the product experience. However, the proposed methodology was limited to cross-modal effects that have temporal difference between prior expectation and posterior experience. We term cross-modal effect with temporal difference *temporal cross-modal effect*. In this respect, the previous methodology was not complete since simultaneous multisensory stimuli also produce cross-modal effects, which we term *simultaneous cross-modal effects*. Researches have been made on this simultaneous cross-modal effect; for instance, Vroomen and De Gelder (2000) revealed that perception of auditory stimuli enhances perceptibility of visual stimuli. Therefore, when we watch a video with synchronized sound, we devote ourselves to the video more than when we watch it without the sound because the presence of the sound increases our sensitivity to the visual information. As this example shows, a simultaneous cross-modal effect also occurs during product experiences.

This paper proposes a methodology to design sensory experiences using a simultaneous cross-modal effect. In the methodology, we extracted users' comprehensive cognitive structure to visualize where the product's features conflict each other, discover cross-modal effects that can help eliminate the conflict, formulate the found cross-modal effects, and discuss design solutions. To prove its effectiveness, the methodology needs to be substantiated. Therefore, we demonstrated the methodology with SLR cameras as a case study of a typical product involving multisensory perception.

## 2 PROPOSED METHODOLOGY

The previous research from our research group (Yanagisawa et al., 2017) proposed a way to discover opportunities to design better product experiences using a temporal cross-modal effect. To update this methodology to be applicable to simultaneous cross-modal effect, we assumed that the methodology would have to match the following requirements. First, the product experience model is structured comprehensively by considering what is perceived and how it is perceived. Second, problems that we could not resolve within a single modality but seemed possible to be solved with a cross-modal effect are extracted. Third, experiments are conducted to formulate the cross-modal effect. Finally, we examine whether the cross-modal effect can be applied to solve the problem and discuss design solutions.

To accomplish the above, we propose a methodology composed of the following five steps:
1. Apply Kansei modeling methodology to the target product and obtain its evaluation structure.
2. Find conflicts where a design element that has a positive effect on the product experience collides with another design element or design limit that has a negative effect on the same experience. Then, explore the cross-modal effect to solve the conflict.
3. Conduct an experiment to obtain responses to the found cross-modal effect by independently manipulating the stimuli from modalities that cause the cross-modal effect.
4. Statistically formulate the cross-modal effect from the result of the experiment.
5. Based on the formulated cross-modal effect, find design solutions for a new product experience.

To discuss the validity of the proposed methodology, we applied above five steps to SLR cameras. The methods for steps 1 and 2 are discussed in detail in 3.1. The application to SLR camera is discussed in Chapter 4. Likewise, the methods for step 3, 4 and 5 are discussed in 3.2. The applications to SLR cameras are in Chapters 5 and 6, which deal with different experiments. Following this, we examine the result of the application and conclude the validity of the proposed methodology.

## 3 METHODS USED IN THE PROPOSED METHODOLOGY

### 3.1 Method to model a cognitive structure and to extract cross-modal opportunities

Yanagisawa et al. (2017) proposed the Kansei modeling methodology for time-series multimodal user experience. In this methodology, they extracted a comprehensive cognitive structure of user Kansei in multisensory interactions between a user and a product. Figure 1 shows an example of the structural model. It can be seen that they placed a series of scenes on the bottom part, and each scene consists of an action-sense pair. Vertically, they set five layers, which are (from top to bottom) delightful experience, delight factor, perceived feature, design element, and scene. Design elements are components of a product that users recognize, and they are connected to each scene. Perceived features are the physical traits that users pay attention to while recognizing design elements, and are connected to each design element. Delight experience is on the top, which is the main goal. Delight factors are placed second from the top originating from delight experience. They are mental stimulations induced by the product to enrich the product experience.

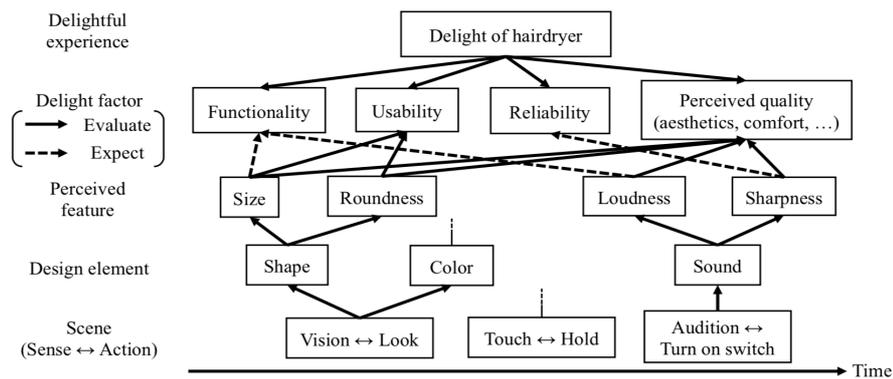

Figure 1. Extraction of cognitive structure of user Kansei

To link delight factors with perceived features, in the previous research, a laddering technique based on the personal-construct theory (Sanui & Maruyama, 1997) was applied. The laddering technique was used to investigate how users' expectations were formed, because the objective of the previous paper was to find a temporal cross-modal effects observed between prior expectation and posterior experience and to explain the cross-modal effect using expectation theory.

In contrast, in this paper, we focus on simultaneous multimodal product experiences. Thus, we simply link delight factors with perceived features with solid/dashed lines if the perceived feature causes positive/negative effect on the delight factor. After developing a structural model of the product, we identified cases wherein one feature improves the delight factor and another feature makes it worse; in such cases, a design conflict may exist, which we termed *conflict*. Then, we discuss if the perception of the features causing the conflict can be changed to resolve the conflict by a cross-modal effect.

### 3.2 Method to formulate a cross-modal effect

To formulate a cross-modal effect, we propose the following procedure. First, construct an experimental environment that can separately manipulate stimuli from two or more modalities that cause a cross-modal effect. Second, conduct the experiment to obtain subjects' responses to the cross-modal effect and evaluate while multisensory stimuli are independently manipulated. In the experiment, the evaluation axis is set according to perceived feature. Finally, analyze the result statistically and achieve a model of the cross-modal effect.

Indeed there are a variety of ways to evaluate a cross-modal effect and statistical methods to model the result; however, in this paper, the evaluation was conducted using three-grade relative evaluation technique and logistic regression analysis was used to formulate the target cross-modal effect. The reason for this selection has been explained in Chapter 5.

## 4 COMPREHENSIVE EXTRACTION OF EVALUATION STRUCTURE OF SLR CAMERA AND CROSS-MODAL OPPORTUNITIES

### 4.1 Method

Based on Kansei modeling methodology, as explained in Chapter 3, we conducted an experiment with 10 participants using five SLR cameras to obtain the evaluation structure. The five cameras were Nikon D7000, OLYMPUS OM-D E-M1 (mirrorless), SONY α7R II (full size mirrorless), Canon EOS 5D Mark II (full size), and CASIO EX-FH25. We selected these cameras with different features, price range, release date, and manufacturer and numbered them 1 to 5. We assumed four scenes such as looking at the camera, holding the camera, focusing, and taking a photo (pushing the shutter button) and performed an interview in each scene separately. We asked participants to compare pairwise two cameras and let them answer which he/she liked better and why. Letting participants compare two cameras would facilitate noticing potential evaluative factors. For example, one would notice that the texture of cameras was important by comparing one that had a rough-treated surface and one that was smooth.

We conducted interviews with both students and camera designers to extract an evaluation structure not only from consumers' viewpoints but also from designers' viewpoints. Also, we asked camera designers about limitations of functions and structures from a manufacturers' point of view. We asked participants to express aloud what they thought and felt as much as possible while evaluating the cameras.

In this experiment, we used varied types of cameras regardless of manufacturer, model, release date, and price, expecting comparisons among various cameras would lead us to comprehensive results. Table 1 shows a list of the SLR cameras used in the experiment and their characteristics.

Ten male volunteers aged between 21 to 50 years participated as experiment evaluators. Six of them were mechanical engineers of cameras and the remaining four were undergraduate or graduate students studying mechanical engineering at the University of Tokyo. The engineers were familiar with the mechanisms of SLR cameras; on the other hand, students were amateurs who only knew how to use it. All participants were physically healthy.

The overall procedure was as follows. We invited the participants into an isolated test room, where they were seated on chairs in front of a table with five SLR cameras placed on it. We asked participants to compare two cameras pairwise and to answer which he/she liked better and which elements of the cameras made him/her feel so in each scene.

For instance, one evaluator answered in the "Shoot" scene: "I prefer number 1. The shutter sound volume of camera number 2 is loud and it made me feel annoyed, while I didn't feel any negative impression from the sound of number 1." From this result, we could extract a hierarchical evaluation structure such as "Sound" (Design Element) originated from "Shoot" (Scene) and engendered "Loudness" (Perceived Feature), and "Loudness" negatively affects "Silence" (Delight Factor) as shown in Figure 2.

### 4.2 Results

Figure 2 shows the results of extracted cognitive structure of SLR cameras' sensory experience. In the figure, the four scenes are lined up in the order of time transition from left to right. Actions and modality of the perception, perceived features, and delight factors in each scene are placed hierarchically. Since the "Shoot" scene is the most fundamental and essential, we thought that the benefit product experience improvements would be maximized by focusing on the "Shoot" scene. Therefore, we focused on the "Shoot" scene and from discovered conflicts we chose the following two items and examined if there is any cross-modal effect that could resolve the conflict.

In following discussions, we emphasized interaction between shutter sounds and shutter vibrations, which were dominant considerations in the "Shoot" scene since there was a high possibility that a cross-modal effect occurs between auditory and tactile stimuli. Many previous studies, e.g., Altinsoy (2010) and Wilson (2010), have shown that multimodality interactions between sound and vibration exist.

1. We defined a crisp sound as a sound that has low reverberation due to its short duration and high attenuation rate. If the sound is crisp, stronger feedback on the act of shooting is obtained and a favorable evaluation is formed, but there was a problem in that the shutter sound is difficult to alter because it is determined by the structure and material of the camera. However, by shifting the lengths of the shutter vibration with respect to that of the shutter sound, it is possible to perceive the shutter sound as shorter or longer on the perceptual level by an audio-tactile cross-modal effect.
2. A shorter lag between the time when the picture is actually taken and the time when the shutter sound is made by pressing the shutter button invites positive feedback from users with regard to shooting action. Meanwhile, the timing of the shutter sound is mechanically lagging behind the time when the picture is actually taken. According to camera engineers, professional photographers, who are amongst the most important customers, frequently point this out. However, by shifting the presentation timing of the shutter vibration, it is possible to perceive the timing of the feedback, which is mechanically restricted, earlier or later at the perceptual level due to the cross-modal effect of auditory and tactile modality.

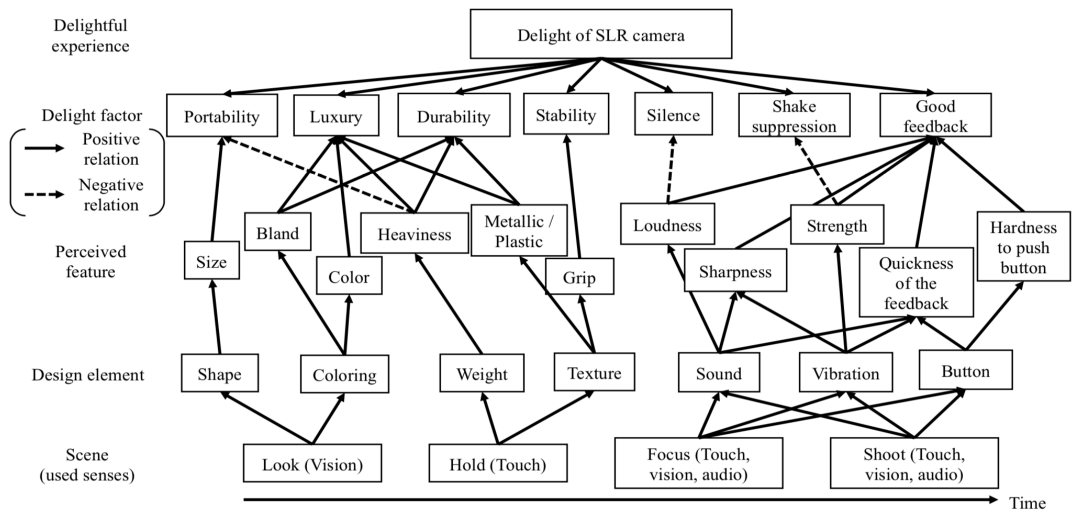

*Figure 2. Extracted cognitive structure of SLR cameras' sensory experience*

### 4.3 Discussion

We considered the above-mentioned two opportunities for cross-modal effects as follows.

For the first opportunity, Gescheider (1967) demonstrated that masking occurs between tactile and auditory modality. Thus, we assumed that the existence of the tactile stimulus dullens the sensitivity to sound because of the masking effect. Given this fact, we hypothesized that the perception of the sound is dulled, and the perceived attenuation rate of the sound is high at the perception level by paying attention to vibration when a vibration with a high attenuation rate is presented at the same time.

The second opportunity was based on the premise that a cross-modal effect will occur if there is a difference in the presentation timing of sound and vibration. In this regard, Spence et al. (1998) showed that, in visual and auditory modality, the presence of a preceding stimulus accelerates the perception timing of the posterior stimulus. Thus, we hypothesized that a similar phenomenon occurred between auditory and tactile senses. At that time, the perceived timing of the camera's feedback will be changed due to the difference between the presentation timing of the shutter sound and shutter vibration.

From the above discussions, since the first cross-modal opportunity was an experiment to handle the attenuation rate and the second opportunity handles the presentation timing, both need to be verified with sudden sound and vibration rather than stationary ones. Furthermore, there were few studies concerning cross-modal effect between sudden sound and vibration. Indeed, Spence et al. (1998) used pulsed tactile stimuli for changing spatial orientation; however, this was focused on spatial changes of perception and it aimed to observe a psychological behavior. The investigation of cross-modal effects between sudden sound and vibration would be useful not only for sensory experience designing of products such as smartphones, gaming devices, and buttons etc., but also for other study fields such as neuroscience and psychology. Therefore, we decided to use sudden stimuli and conducted experiments on whether the perception of the length of composite feedback of sound and vibration varies by cross-modal effect when changing the combination of sound and vibration length (Experiment 1) and whether the perception of the presentation timing of composite feedback varies by cross-modal effect when changing the combination of presentation timing of sound and vibration stimulus (Experiment 2). As the above practice shows, extracting and visualizing users' cognitive structure by arranging the evaluative structure hierarchically was effective in finding design conflicts. We first extracted conflicts comprehensively and second, investigated cross-modal effects as solutions to the found conflicts. This methodology is beneficial for conflict extraction; however, identification of possible cross-modal effects depends on the experimenter's idea to some extent, which is a limitation of the methodology.

# 5 EXPERIMENT 1: EFFECT OF DECAY TIME DIFFERENCE BETWEEN AUDITORY AND TACTILE STIMULI ON PERCEPTION OF RESPONSE ATTENUATION

## 5.1 Method

The hypothesis of this experiment was that the perceived attenuation time of composite feedback of sound and tactile stimuli becomes shorter when the attenuation time of the tactile stimulus is shortened. "Length" refers to the length of the attenuation time when there is no special explanation.

We set a pair composed of sound and tactile stimulus, both of which have a length of 120 msec, as a control and investigated whether the perception of the composite feedback length changes when compared with a pair whose length of tactile stimulus is shorter or longer than that of the control.

In Figure 3, the pair shown in the center is the control. We manipulated the length of the tactile stimulus to 40, 60, 80, 100, 120, 150, 180, 210, and 240 msec respectively. The one shown on the left of the figure is the pair of 120 msec sound and 40 msec (shortest) vibration, and the right one shows the pair of 120 msec sound and 240 msec (longest) vibration. In this way, we created nine pairs.

Fifteen male and female volunteers aged 20 to 25 years participated the experiment. They were all physically healthy undergraduate or graduate students studying mechanical engineering at the University of Tokyo.

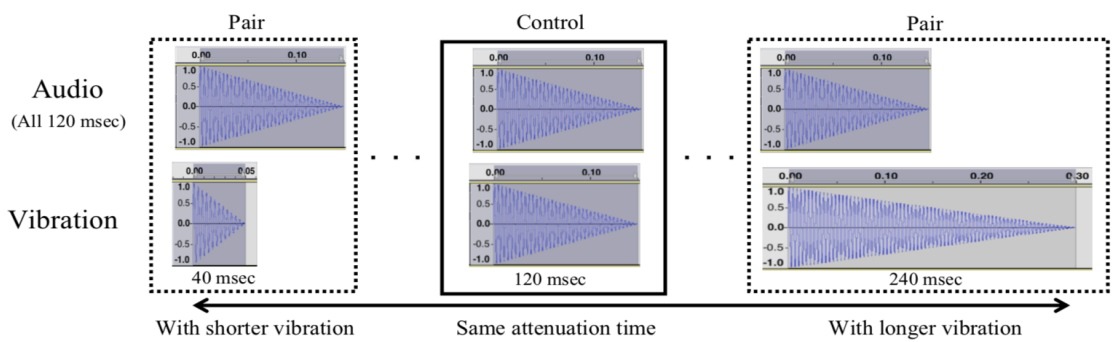

*Figure 3. Stimuli pairs for Experiment 1*

We asked participants to evaluate the samples by grading whether the length of the samples was perceived as "Shorter," "Same," or "Longer" compared to the reference. Each sample was presented five times in total in a random order. (In other words, participants did 45 evaluations in total.)

We set these three choices to perform a logistic regression analysis. Since participants could not distinguish "Much shorter," "Shorter," and "Slightly shorter" in preliminary experiment, we considered it difficult to perform the semantic differential method. Thus, logistic regression analysis was selected to achieve a model of the target cross-modal's behavior. To perform a logistic regression analysis, we needed an odds ratio. For this ratio, we used the following value: probability of answering "Shorter" divided by the probability of answering "Not shorter." "Not shorter" is equal to the sum of probability of answering "Same" and "Longer." We used the three-step evaluation to perform a logistic regression analysis not only to conduct "Shorter/Not shorter" analysis but also to perform "Longer/Not longer" analysis by likewise treating the sum of "Shorter" and "Same" as "Not longer". In this paper, we do not show the result of "Longer/Not longer" analysis, since this result did not lead us to a design solution; however, it was academically significant to model the behavior of **longer side.** Experiment 2 in **Chapter 6** was also conducted with three-step grading for the same reason.

For stimuli selection, Wilson (2010) reported that the effect of cross-modality between auditory and tactile stimuli is maximized when both frequencies are equal. Though this result was for full body stimuli, we assumed this fact could also be applied to fingertips. Additionally, Verrillo (1971) noted that the sensitivity to vibration becomes highest at the frequency of 250 Hz due to Pacinian corpuscles. Therefore, in order to maximize the sensitivity to stimuli, we used both the sound and vibration stimuli with a frequency of 250 Hz. For the length of the stimuli, we found that the shutter sound of SLR cameras can be roughly divided into two parts: i) front curtain and charge noise and ii) second curtain noise. Both parts were approximately 120 msec. Thus, we concluded experiments with stimuli of 120 msec. Also, to simplify the experiment, both stimuli had simple linear attenuation waveform. In Experiment 1, sounds and vibrations were presented simultaneously and the lengths of the tactile stimuli were manipulated, whereas the length of the sound stimuli was fixed at 120 msec.

We created an experimental system that consists of a computer, two microcomputers (Arduino Uno), a shaker (Brüel & Kjaer 4810), an amplifier (Brüel & Kjaer 2706) connected to the shaker, and earphones (Bose QuietComfort 20). Two microcomputers were connected to the computer and one was connected to the earphones to present auditory stimuli, whereas the other was connected to the shaker through the amplifier to present tactile stimuli to the fingertip. In this system, a button was utilized, and we could present stimuli by pressing it. In Experiment 1, an experimenter pressed the button. Participants wore ear-muffs (Peltor H540A) so that the sounds from the shaker would not interfere.

Fifteen male and female volunteers aged 21 to 25 years participated. They were physically healthy undergraduate or graduate students studying mechanical engineering at the University of Tokyo.

## 5.2 Result

We converted the result into a two-step evaluation of "Shorter" and "Not shorter" by totaling the number of people who answered "Same" and "Longer" and treating it as "Not shorter." Then, we obtained the probability of answering "Shorter" and "Not shorter," and conducted a logistic regression analysis with this probability as an objective variable and the length of tactile stimulus as an explanatory variable. The odds ratio can be calculated by dividing probability of answering "Shorter" by "Not shorter."

Figure 4 shows the result. In the figure, the circles plotted on the graph are the raw data values, and the solid line curve is the obtained regression curve. The broken lines show the 5% confidence zone of the odds ratio. If a plot is in the area between the broken lines, the regression model is significant at the risk ratio of 5% around the plot. From the model, probability of answering "Shorter" is correlated with the length of tactile stimulus. When the vibration is short, the probability is low, whereas when the vibration is long, the probability is high. The plots of 120, 150, and 240 msec are outside the confidence area, and so the regression model is significant at 5% risk ratio in the range of 100 msec or less. However, around 120, 150, and 240 msec, the model is not reliable enough.

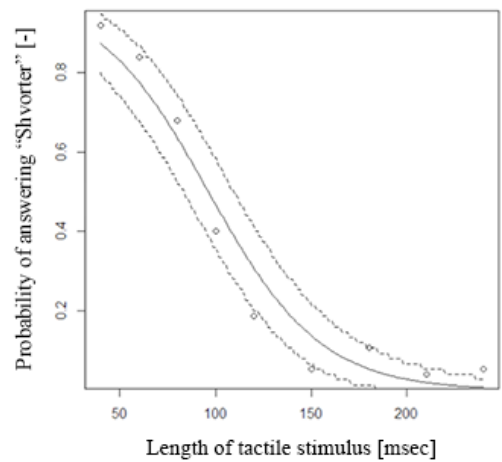

*Figure 4. Regression curve with confident zone of Experiment 1. Horizontal axis: length of tactile stimulus [msec], vertical axis: probability of responding "Shorter" [-].*

## 5.3 Discussion

From the result, for composite feedback of auditory and tactile stimuli, the participants perceived it "Shorter" when the length of tactile stimulus is shortened, while the participants perceived "Longer" when it is lengthened. This result suggests that by decreasing only the attenuation time of shutter vibration, the perception of the attenuation time of the whole shutter feedback will be perceived shorter, which follows our hypothesis.

Even if the shutter sound is established as a brand image of the manufacturer and is difficult to change, users will perceive the whole feedback as shorter and crisper than the original one if the shutter vibration is modified to shorter one. Therefore, for example, it is possible to design a better product experience by adding a shorter sudden vibration and making the feedback crisper at perceptual level without changing the shutter sound.

# 6 EXPERIMENT 2: EFFECT OF THE TEMPORAL GAP BETWEEN AUDITORY AND TACTILE STIMULI ON PERCEPTION OF RESPONSE TIMING

## 6.1 Method

The hypothesis of this experiment was that the perceived presentation timing of composite feedback of sound and tactile stimuli hastened when the presentation timing of the tactile stimulus came earlier. Difference of timings means the difference of presentation timings of auditory and tactile stimuli if there is no special explanation.

We set a pair with simultaneous auditory and tactile stimulus as a control and investigated whether the perception of the presentation timing of the composite feedback changes when this control is compared with a pair whose tactile stimulus is presented earlier or later than auditory one.

In Figure 5, the simultaneous pair shown in the center was the control. We manipulated the presentation timing of the tactile stimulus to -100, -80, -60, -40, -20, 0, 10, 20, 30, 40, and 50 msec later than that of the auditory ones respectively. Negative values indicate vibration was presented earlier than sound. The one shown on the left of the figure is the pair whose tactile stimulus was 100 msec earlier than auditory one, whereas the right one is the pair whose tactile stimulus was 50 msec later. In this way, we created eleven pairs.

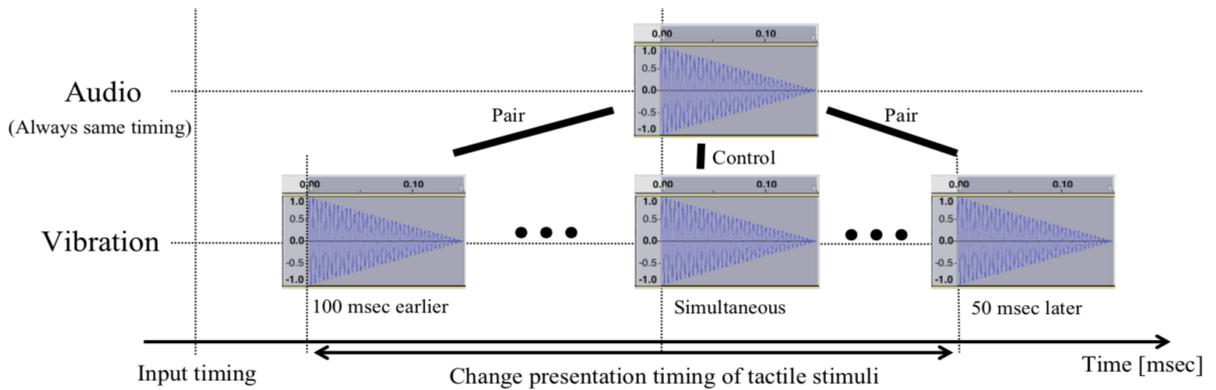

*Figure 5. Stimuli pairs for Experiment 2*

Similar to Experiment 1, we asked participants to evaluate the samples by grading whether the presentation timing of the sample was perceived as "Earlier," "Same," or "Later" compared to the reference. Each sample was presented five times in total in random order. The attenuation time of both stimuli were fixed to 120 msec and presentation timing of the auditory stimulus was fixed. All properties other than these were same as those we used in Experiment 1.

The experiment environment and participants were also the same as Experiment 1 except the way to present stimuli: we asked participants to press the button by themselves to present stimuli instead of an experimenter pressing it.

## 6.2 Result

Similar to Experiment 1, by totalizing the number of people who answered "Same" and "Later" and treating it as "Not earlier," we converted it into a two-step evaluation of "Earlier" and "Not earlier." Then, we obtained the probability of answering "Earlier" and conducted a logistic regression analysis with the probability as the objective variable and the difference of timings as the explanatory variable. Figure 6 shows the result. In the figure, style properties other than axes are same as Figure 4. From the model, probability of answering "Earlier" decreases along with an increase of the difference of timings. Since every plot is inside the confidence area, the regression model is significant at 5% risk ratio everywhere.

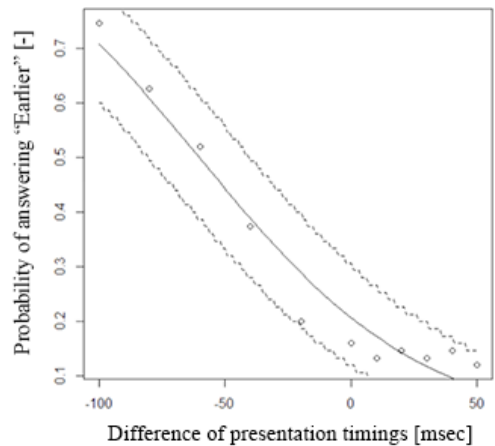

*Figure 6. Regression curve with confident zone for Experiment 2. Horizontal axis: difference of presentation timings [msec], vertical axis: probability of answering "Earlier" [-].*

## 6.3 Discussion

From the result, for composite feedback of auditory and tactile stimuli, the participants perceived it "Earlier" when the presentation timing of vibration was earlier, while the participants perceived "Later" when the vibration presentation timing was delayed. This result suggests that, by bringing the shutter vibration forward, the perception of the presentation timing of whole shutter feedback is perceived earlier, which supports our hypothesis.

Even if the timing of the shutter sound mechanically lags behind the actual time the photo is captured, by shifting the timing of shutter vibration, users will perceive the whole feedback earlier. Therefore, for example, by adding vibration that is presented earlier than the shutter, it is possible to design a better product experience by shifting the timing of the feedback closer to the real timing of photo capture at the perceptual level.

## 7 CONCLUSION

We proposed a methodology to extract opportunities of cross-modal effects, formulate the target cross-modal effect, and apply it to improve sensory product experience. In the methodology, we updated the Kansei modeling methodology to be used to extract simultaneous cross-modal opportunities by changing the technique to link perceived features with delight factors. Furthermore, we described how to formulate simultaneous cross-modal effects and how to collect data by subjective evaluations.

We applied the methodology to find a solution that improves the multisensory experience of SLR cameras by using a cross-modal effect. From the case study, we found two solutions: 1) a vibration with shorter attenuation time makes the perceived composite feedback of sound and vibration crisper, and 2) a vibration presented earlier makes the perceived presentation timing of composite feedback earlier. We formalized the two cross-modal effects in a quantitative manner for application of these formulations to other product experience design as scientific knowledge. The fact that the methodology led us to solutions of sensory SLR camera experiences substantiated that the proposed methodology helps to 1) discover design conflicts comprehensively and investigate opportunities of simultaneous cross-modal effects to resolve the conflicts, and 2) analyze the target cross-modal effect statistically and obtain a regression model for solution finding. Therefore, we conclude that the proposed methodology provides an effective way to design a sensory product experience using simultaneous cross-modal effects.

## REFERENCES

Altinsoy, M. E., and Merchel, S. (2010). "Cross-modal frequency matching: Sound and whole-body vibration", *Lecture Notes in Computer Science* (including subseries *Lecture Notes in Artificial Intelligence and Lecture Notes in Bioinformatics*), https://doi.org/10.1007/978-3-642-15841-4_5

Charpentier, A. (1891). "Analyse experimentale: De quelques elements de la sensation de poids. [Experimental analysis: On some of the elements of sensations of weight] ", *Archives de Physiologie Normale et Pathologique*, 3, pp. 122–135.

Deliza, R., and MacFie, H. J. H. (1996). "The generation of sensory expectation by external cues and its effect on sensory perception and hedonic ratings: A review", *Journal of Sensory Studies*, 11(2), pp. 103–128.

Driver, J., and Spence, C. (1998). Cross-modal links in spatial attention, *Philosophical Transactions of the Royal Society of London*, Series B, Biological Sciences, 353(1373), pp. 1319–31. https://doi.org/10.1098/rstb.1998.0286wi

Gescheider, G. A., and Niblette, R. K. (1967). "Cross-modality masking for touch and hearing", *Journal of Experimental Psychology*, 74(3), pp. 313–320. https://doi.org/10.1037/h0024700.

Klatzky, R. L., Lederman, S. J., and Matula, D. E. (1993). "Haptic exploration in the presence of vision", *Journal of Experimental Psychology: Human Perception and Performance,* 19(4), pp. 726–743.

Lederman, S. J., Thorne, G., and Jones, B. (1986). "Perception of texture by vision and touch: Multidimensionality and intersensory integration", *Journal of Experimental Psychology: Human Perception and Performance*, 12(2), pp. 169–180.

Ludden, G., Schifferstein, H., and Hekkert, P. (2009). "Visual-tactual incongruities in products as sources of surprise", *Empirical Studies of the Arts,* 27(1), pp. 61–87.

Schifferstein, H. N. J., Fenko, A., Desmet, P. M. A., Labbe, D., and Martin, N. (2013). "Influence of package design on the dynamics of multisensory and emotional food experience", *Food Quality and Preference*, 27(1), pp. 18–25. https://doi.org/10.1016/j.foodqual.2012.06.003

Spence, C., Nicholls, M. E. R., Gillespie, N., and Driver, J. (1998). "Cross-modal links in exogenous covert spatial orienting between touch, audition, and vision", *Perception and Psychophysics*, 60(4), pp. 544–557. https://doi.org/10.3758/BF03206045

Verrillo, R. T. (1971). "Vibrotactile thresholds measured at the finger", *Perception & Psychophysics,* 9(4), pp. 329–330.

Vroomen, J., and De Gelder, B. (2000). "Sound enhances visual perception: Cross-modal effects of auditory organization on vision", *Journal of Experimental Psychology: Human Perception and Performance*, 26(5), pp. 1583–1590. doi: 10.1037/0096-1523.26.5.1583.

Wilson, E. C., Reed, C. M., and Braida, L. D. (2010). "Integration of auditory and vibrotactile stimuli: Effects of frequency", *The Journal of the Acoustical Society of America*, 127(5), pp. 1960-1974. https://doi.org/10.1121/1.3365318

Yanagisawa, H., Miyazaki, C, and Bouchard, C. (2017). "Kansei modeling methodology for multisensory UX design", *International Conference on Engineering Design (ICED17)*, Vancouver, Canada, 21–25 August 2017, Vol. 8: Human Behaviour in Design, pp. 159–168.

Yanagisawa, H., Miyazaki, C. and Nakano, S. (2016), "Kansei modeling for multimodal user experience (visual expectation effect on product sound perception) ", *Proceedings of the INTER-NOISE 2016 - 45th International Congress and Exposition on Noise Control Engineering: Towards a Quieter Future*, pp. 1793–1802.

Yanagisawa, H., and Takatsuji, K. (2015). "Effects of visual expectation on perceived tactile perception: An evaluation method of surface texture with expectation effect", *International Journal of Design*, 9(1), pp. 39–51.

**ACKNOWLEDGMENTS**

Part of this work was supported by JSPS KAKENHI Grant Number 18H03318 and Sony Corporation. We would like to thank Professor Tamotsu Murakami, Dr. Kazutaka Ueda of The University of Tokyo, Dr. Kazuko Yamagishi of Sony, and members of the Design Engineering Laboratory at The University of Tokyo for supporting this project.